\documentclass[proceedings, preprint]{rmaa}

\usepackage{psfrag,color}
\usepackage{amsmath}

\SetYear{2013}
\SetConfTitle{Magnetic Fields in the Universe IV}

\title{Smoothed Particle Magnetohydrodynamics Simulations of Protostellar Jets and Turbulent Dynamos} 

\author{
  T. S. Tricco,\altaffilmark{1} 
  D. J. Price,\altaffilmark{1}
  C. Federrath,\altaffilmark{1}
  and M. R. Bate \altaffilmark{2}}

\altaffiltext{1}{Monash Centre for Astrophysics,
  School of Mathematical Sciences,
  Monash University,
  Vic, 3800, Australia
  (terrence.tricco@monash.edu, daniel.price@monash.edu, christoph.federrath@monash.edu).}

\altaffiltext{2}{School of Physics and Astronomy,
  University of Exeter,
  Stocker Road, Exeter,
  EX4 4QL, United Kingdom
  (mbate@astro.ex.ac.uk)}

\shortauthor{Tricco, et al}
\shorttitle{SPMHD Simulatiions of Protostellar Jets and Turbulent Dynamos}

\listofauthors{T. S. Tricco, D. J. Price, C. Federrath, \& M. R. Bate}
\indexauthor{Tricco, T. S.}
\indexauthor{Price, D. J.}
\indexauthor{Federrath, C.}
\indexauthor{Bate, M. R.}

\abstract{We presents results from Smoothed Particle Magnetohydrodynamics simulations of collapsing molecular cloud cores, and dynamo amplification of the magnetic field in the presence of Mach 10 magnetised turbulence.  Our star formation simulations have produced, for the first time ever, highly collimated magnetised protostellar jets from the first hydrostatic core phase.  Up to 40\% of the initial core mass may be ejected through this outflow.  The primary difficulty in performing these simulations is maintaining the divergence free constraint of the magnetic field, and to address this issue, we have developed a new divergence cleaning method which has allowed us to stably follow the evolution of these protostellar jets for long periods.  The simulations performed of supersonic MHD turbulence are able to exponentially amplify magnetic energy by up to 10 orders of magnitude via turbulent dynamo.  To reduce numerical dissipation, a new shock detection algorithm is utilised which is able to track magnetic shocks throughout a large range of magnetic field strengths.}

\resumen{}

\addkeyword{stars: formation}
\addkeyword{ISM: Jets and outflows}
\addkeyword{ISM: magnetic fields}
\addkeyword{turbulence}
\addkeyword{methods: numerical}

\begin{document}
\maketitle

\section{Introduction}
\label{sec:intro}

Magnetic fields play an important role in all phases of star formation. On the scale of molecular clouds, the sites where stars are formed, magnetic fields can affect the structure of the supersonic turbulence.  It is known that this turbulence drives the local compression needed to trigger gravitational collapse of molecular cloud cores, and simulations by \citet{pn11} and \citet{fk12} show that the additional pressure from magnetic fields can help support against gravitational collapse and reduce star formation rates by factors of 2--3.  

On the scale of individual protostars, magnetic fields are responsible for launching protostellar jets and outflows.  These jets and outflows are both well observed \citep{richeretal00,wuetal04} and simulated \citep[e.g.][]{mim08, commerconetal10, ptb12, mh13}.  They reduce the efficiency of star formation \citep{mm00,hansenetal12}, and are drivers of turbulence in the interstellar medium \citep{nl07,cfb10}.

In this work, we have performed Smoothed Particle Magnetohydrodynamic (SPMHD) \citep{pm04a,pm04b,pm05} simulations of the gravitational collapse of a prestellar core to form the first hydrostatic core (before the protostar is formed), and also of magnetised Mach 10 turbulence which exponentially amplifies an initial seed magnetic field.   These simulations use new numerical techniques to maintain the divergence free constraint on the magnetic field, detect shocks, and reduce numerical dissipation of the magnetic field.

This paper begins with a brief review of SPMHD in \S\ref{sec:spmhd}, discussing how magnetic fields are simulated with SPMHD, and the benefits of using SPMHD for star formation simulations.  In \S\ref{sec:cleaning}, the constrained hyperbolic divergence cleaning method is introduced, which is used to maintain $\nabla \cdot {\bf B}=0$.  In \S\ref{sec:jet}, we present results from simulations of a collapsing prestellar core.  
The jet produced in our simulations during protostellar collapse is discussed in \S\ref{sec:jet-results}.  In \S\ref{sec:mhdturb}, results from simulations of Mach 10 magnetised turbulence are presented.  These simulations focus on the dynamo amplification of the magnetic field, with comparison to results from grid based methods.  A summary and discussion is given in \S\ref{sec:summary}.

\section{Smoothed Particle Magnetohydrodynamics}
\label{sec:spmhd}

Smoothed Particle Hydrodynamics (SPH) is a numerical method for simulating fluid flow (see reviews by \citealt{monaghan05,price12}).  The hydrodynamic equations are solved by discretising the fluid into a set of particles which contain a portion of the mass, energy, and momentum of the fluid.  Fluid quantities, such as density, are calculated per particle by interpolating from neighbouring particles using a kernel weighted summation. 

SPH is widely used in astrophysics.  It can easily handle complex geometries, has excellent conservation properties, and couples easily with N-body methods for gravity.  Regions of higher density contain more mass, and therefore more resolution elements.  This is useful for star formation, because as gas collapses to form dense objects, the mass, and hence particles, trace the collapse providing continuous resolution of that process until the Jeans mass falls below the mass resolution \citep{bb97}.

\subsection{The equations of SPMHD}
The ideal MHD equations solved with SPMHD are given, for a particle $a$, by
\begin{align} 
\rho_a &=  \sum_b m_b W_{ab} (h_a), \label{eq:sphcty} \\
h_{a} &= h_{\rm fac} \left( \frac{m_{a}}{\rho_{a}}\right)^{1/n_\text{dim}}, \label{eq:h} \displaybreak[0] \\
\frac{{\rm d}{\bf{v}}_a}{{\rm d}t} &= - \sum_b m_b \bigg[\frac{{\bf M}_a}{\Omega_a \rho_a^2}\cdot \nabla_a W_{ab}(h_a) \nonumber \\
& \hspace{16mm}+ \frac{{\bf M}_{b}}{\Omega_b \rho_b^2} \cdot \nabla_a W_{ab}(h_b) \bigg], \label{eq:spmhd-momentum-eqn} \displaybreak[0] \\
 \frac{{\rm d}{\bf{B}}_a}{{\rm d}t} &= - \frac{1}{\Omega_a \rho_a} \sum_b m_b \bigg[ {\bf{v}}_{ab} \left( {\bf{B}}_a \cdot \nabla_a W_{ab}(h_a) \right) \nonumber \\
& \hspace{20mm}- {\bf{B}}_a \left( {\bf{v}}_{ab} \cdot \nabla_a W_{ab}(h_a) \right) \bigg], \label{eq:sphind}
\end{align}
where ${\bf v}$ and ${\bf B}$ are the velocity and magnetic fields, ${\bf v}_{ab}$ denotes ${\bf v}_a - {\bf v}_b$, and $W$ is an interpolation kernel.  We use the cubic spline kernel \citep{ml85} in this work. 

Variable resolution is obtained by self-consistently deriving the density, $\rho$, and smoothing length, $h$, through iteration of equations (\ref{eq:sphcty}) and (\ref{eq:h}).  The smoothing length is related to the local particle spacing by $h_{\rm fac}=1.2$, with $n_\text{dim}$ corresponding to the number of dimensions.  Variable smoothing length gradients are handled by $\Omega$ (see \citealt{sh02}).  

The momentum equation (\ref{eq:spmhd-momentum-eqn}) is derived from the Lagrangian and exactly conserves energy and momentum.  It is expressed in terms of the Maxwell stress tensor,
\begin{equation}
{\bf M} = \frac{{\bf B}{\bf B}}{\mu_0} - \left(P + \frac{B^2}{2\mu_0} \right) {\bf I} ,
\label{eq:divM}
\end{equation}
with the thermal pressure, $P$, obtained through a suitable equation of state.  Terms containing $\nabla \cdot {\bf B}$ are subtracted from the SPMHD momentum equation, introducing a small amount of non-conservation of energy and momentum, but greatly enhancing stability and performance.  The induction equation (\ref{eq:sphind}) is a representation of ${\rm d}{\bf B}/{\rm d} t = -{\bf B}(\nabla \cdot {\bf v}) + ({\bf B} \cdot \nabla) {\bf v}$.

\subsection{Shock capturing}
\label{sec:spmhddiss}
Hydrodynamic and magnetic shocks are captured by the addition of artificial viscosity and resistivity to the momentum and induction equations.  The artificial viscosity used here was formulated by \citet{monaghan97} by analogy to Riemann solvers, and is given by
\begin{equation}
 \left( \frac{{\rm d}{\bf v}_a}{{\rm d}t} \right)_\text{diss} = \sum_b m_b \frac{\alpha v_\text{sig}}{\overline{\rho}_{ab}} {\bf v}_{ab} \cdot \hat{{\bf r}}_{ab} \nabla_a W_{ab} .
\label{eq:visc}
\end{equation}
The signal velocity represents the characteristic speed of information propagation across the shock, given by
\begin{equation}
 v_\text{sig} = 0.5 \left( c_a + c_b - \beta {\bf v}_{ab} \cdot \hat{{\bf r}}_{ab} \right) .
\label{eq:vsig}
\end{equation}
For ideal MHD, the sound speed, $c$, is replaced by the fast MHD wave speed,
\begin{align}
 v =& \frac{1}{\sqrt{2}} \bigg[ \left(c^2 + v_A^2\right) \nonumber \\
&\hspace{5mm}+ \left[ (c^2 + v_A^2)^2 - 4 c^2 v_A^2 (\hat{{\bf B}} \cdot \hat{{\bf r}}_{ij}) \right]^{1/2} \bigg]^{1/2} .
\label{eq:vsigfastmhd}
\end{align}
where $v_A$ corresponds to the Alfv\'en speed.

Artificial resistivity was formulated through a similar procedure by \citet{pm05}.  The corresponding term in the induction equation (\ref{eq:sphind}) is given by,
\begin{equation}
 \left( \frac{{\rm d}{\bf B}_a}{{\rm d}t} \right)_\text{diss} = \rho_a \sum_b m_b \frac{\alpha_B v^{B}_\text{sig}}{\overline{\rho}_{ab}^2} \left({\bf B}_a - {\bf B}_b\right) \hat{\bf{r}}_{ab} \cdot \nabla_a W_{ab} .
\label{eq:resistivity}
\end{equation}
The signal velocity for artificial resistivity is chosen as $v_\text{sig}^B= 0.5 (v_a + v_b)$, which is the averaged fast MHD wave speeds. 

The dimensionless parameters $\alpha$ and $\alpha_B$ are of order unity.  To reduce the dissipation from artificial viscosity and resistivity away from shocks (where it is unnecessary), $\alpha$ and $\alpha_B$ may be set individual for each particle and used to regulate the strength of the applied dissipation.  \citet{mm97} proposed integrating $\alpha_a$ according to 
\begin{equation}
 \frac{{\rm d} \alpha_a}{{\rm d}t} = \max(- \nabla \cdot {\bf v}_a, 0) - \frac{\alpha_a - \alpha_\text{min}}{\tau} ,
\label{eq:intalpha}
\end{equation}
with $\alpha_a \in [0.1,1]$.  This equation increases $\alpha_a$ in regions of converging flow, with a post-shock decay timescale, $\tau = h / C c$, of approximately five smoothing lengths ($C\sim0.1$).

\citet{pm05} created a similar switch for artificial resistivity, using
\begin{equation}
 \frac{{\rm d} \alpha_{B,a}}{{\rm d}t} = \max(\vert \nabla \times {\bf B}_a \vert, \vert \nabla \cdot {\bf B}_a \vert) - \frac{\alpha_{B,a} - \alpha_{B,\text{min}}}{\tau} ,
\label{eq:intalphaB}
\end{equation}
with a range $\alpha_{B,a}\in[0,1]$.

Recently, we have proposed a new switch for artificial resistivity that is more robust at detecting shocks and leads to less overall dissipation \citep{tp13}.  It sets
\begin{equation}
 \alpha_{B,a} = \frac{h \vert \nabla {\bf B}_a \vert}{\vert {\bf B}_a \vert} ,
\end{equation}
in the range $\alpha_{B,a}\in[0,1]$.  This increases artificial resistivity in regions of strong magnetic field gradients.  Since $v_A \propto B$, this leads to a quantity which is related to the Alfv\'enic Mach number.  By normalising the gradient against the magnitude of the magnetic field, the switch responds to the relative degree of discontinuity and does so independently of the absolute magnetic field strength (this is important for the dynamo amplification simulations in \S\ref{sec:mhdturb}).   Setting the value of $\alpha_B$ directly in this manner improves the responsiveness to shocks by removing the time delay present in equation (\ref{eq:intalphaB}).

\section{Constrained hyperbolic divergence cleaning}
\label{sec:cleaning}

The zero divergence constraint on the magnetic field is maintained using constrained hyperbolic divergence cleaning \citep{tp12}.  The cleaning algorithm couples an additional scalar field, $\psi$, to the magnetic field, and divergence error in the magnetic field is dispersed and diffused using a series of damped waves.  By spreading the divergence error over a larger volume, it is able to be removed faster than using just a diffusion term alone, and furthermore, the impact of any single large source of error is reduced.

The method is a Hamiltonian version of hyperbolic divergence cleaning \citep{dedner02} which has been derived by defining the energy content of the $\psi$ field, and including it as part of the Lagrangian.  This leads to continuum equations,
\begin{align}
 \frac{{\rm d}{\bf B}}{{\rm d}t} =& - \nabla \psi , \\
\frac{{\rm d}\psi}{{\rm d}t} =& - c_h^2 \nabla \cdot {\bf B} - \frac{\psi}{\tau} - \tfrac{1}{2} \psi \left(\nabla \cdot {\bf v}\right) , \label{eq:continuumcleandpsi}
\end{align}
and SPMHD equations
\begin{align} 
\left( \frac{{\rm d}{\bf B}_a}{{\rm d}t} \right)_\psi =& -\rho_a \sum_b m_b \bigg[ \frac{\psi_a}{\Omega_a \rho_a^2} \nabla_a W_{ab}(h_a) \nonumber \\
&\hspace{16.5mm}+ \frac{\psi_b}{\Omega_b \rho_b^2} \nabla_a W_{ab}(h_b) \bigg], \label{eq:sphgradpsi}\\
\frac{d\psi_{a}}{{\rm d}t} =& \frac{c_h^2}{\Omega_a \rho_a} \sum_b m_b {\bf B}_{ab} \cdot \nabla_a W_{ab}(h_a) \nonumber \\
&- \frac{\psi_a}{\tau} \nonumber \\
&+ \frac{\psi_a}{2 \Omega_a \rho_a} \sum_b m_b {\bf v}_{ab} \cdot \nabla_a W_{ab}(h_a) . \label{eq:sphcleandpsi}
\end{align}
The divergence wave speed, $c_h$, is chosen as the maximum allowable according to the Courant timestep criterion (typically this would be the fast MHD wave speed).  The damping term, $\tau = h / \sigma c_h$, is best chosen in the regime of critical damping, which from empirical tests \citep{tp12}, is $\sigma \in [0.2,0.3]$ for 2D and $\sigma \in [0.8,1.2]$ for 3D.   Equation (\ref{eq:continuumcleandpsi}) differs from \citet{dedner02} by the addition of the $- \tfrac{1}{2} \psi \left(\nabla \cdot {\bf v}\right)$ term, which describes how $\psi$ changes as the fluid is expanded or compressed.  The constraint from energy conservation also leads to a specific choice of derivative operators in the SPH formulation, as given by equations (\ref{eq:sphgradpsi}) and (\ref{eq:sphcleandpsi}).

By building the method from the ground up in the context of the Lagrangian equations of motion, it inherently retains the stability properties of SPH and is guaranteed to always decrease the divergence of the magnetic field.  This fixes problems in the previous implementation by \citet{pm05} particularly at density contrasts and free boundaries.

\section{Molecular cloud core collapse}
\label{sec:jet}

\citet{larson69} showed collapsing prestellar cores undergo a two stage collapse for low mass star formation.  For both stages the collapse is nearly isothermal, but has a brief adiabatic phase when molecular hydrogen reaches densities higher than $10^{-13}$~g~${\rm cm}^{-3}$.  At these densities, the gas becomes optically thick, trapping radiation.  These first hydrostatic core objects have a lifetime of only several thousand years, ending when the gas reaches $2000$~K and the molecular hydrogen disassociates.  The radiation is then able to freely escape, and the core undergoes a second near-isothermal collapse to form a protostellar core.

Observations of first core objects have been difficult because of the low luminosity of these objects, and it is only within recent years that candidate detections have been made.  \citet{pinedaetal11} observed a low-mass dense core in the Perseus Molecular Cloud, with upper limits on bolometric luminosity and temperature of $0.05$~L$_\odot$ and $30$~K, and found traces of a slow ($3$ km ${\rm s}^{-1}$), poorly collimated outflow.  Per Bolo 58 has been studied by \citet{enochetal10}, finding it to be a promising first hydrostatic core candidate with an internal luminosity of $\sim 0.01 L_\odot$.  \citet{dunhametal11} detected a well collimated ($\sim8^\circ$) bipolar outflow in Per Bolo 58 with characteristic velocity of $2.9$ km ${\rm s}^{-1}$.

We have performed simulations of a $1M_\odot$ collapsing prestellar core during the first stage of collapse to form the first hydrostatic core \citep*{ptb12}.  

\subsection{Initial conditions and numerical details}

\begin{figure}
 \includegraphics[width=\linewidth]{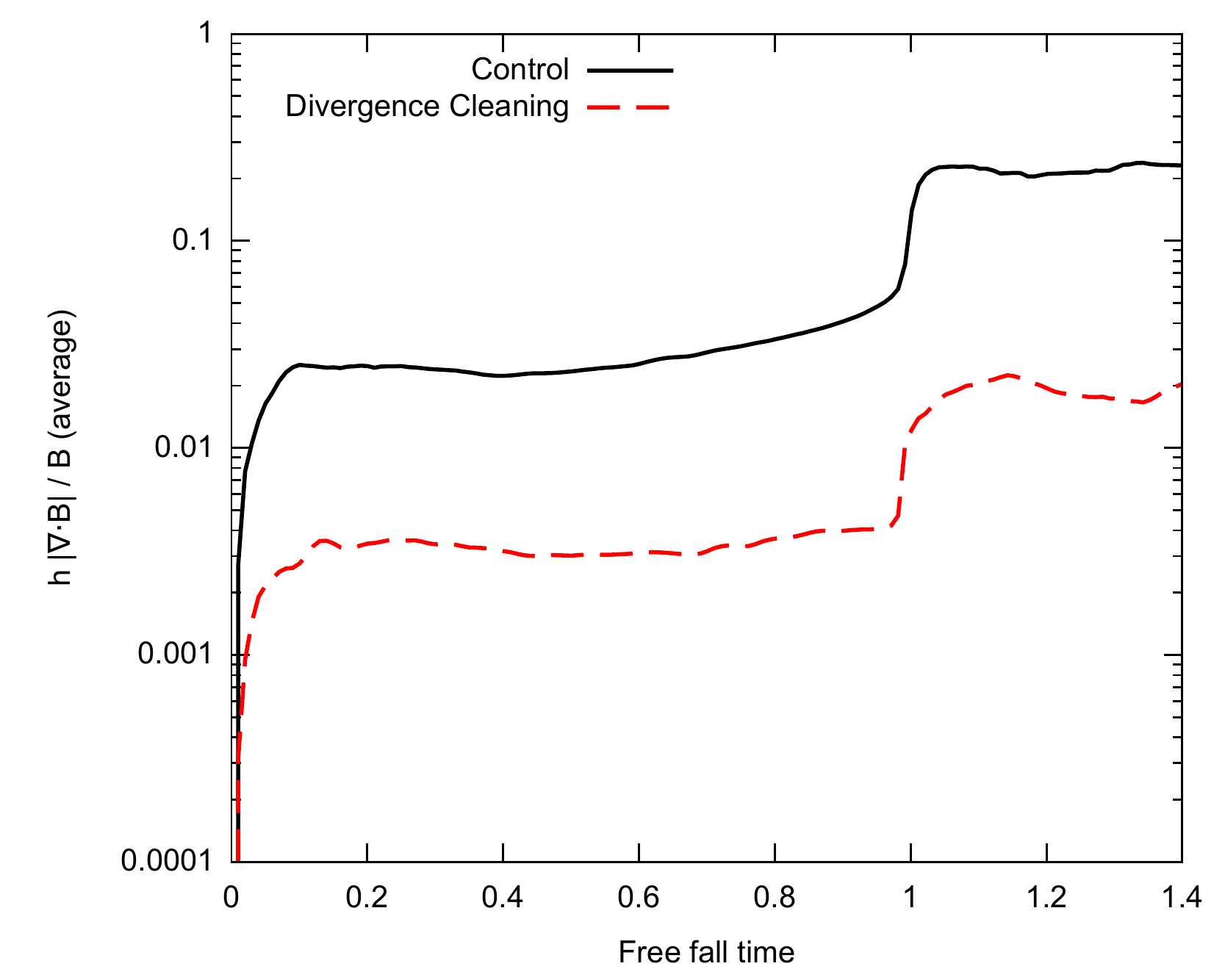}
\caption{Average divergence error, $h \vert \nabla \cdot {\bf B} \vert / \vert {\bf B} \vert$, of the gravitational collapse simulation with (red, dashed line) and without (black, solid line) the constrained hyperbolic divergence cleaning method.  When cleaning is applied, the average divergence error of the magnetic field is reduced by an order of magnitude and kept within 1\%.}
\label{fig:jet-cleaning}
\end{figure}

The simulations are performed for a $1M_\odot$ spherical core with radius $R=4 \times 10^{16}$~cm ($\sim2700$ AU), giving an initial density of $7.43 \times 10^{-18}$~g~${\rm cm}^{-3}$.  It is set in solid body rotation with angular velocity $\Omega = 1.77 \times 10^{-13}$~rad~${\rm s}^{-1}$.  The free fall time is $t_\text{ff}\simeq 24000$~yr.  A barotropic equation of state is used (as described in \citealt{ptb12}), where the gas is isothermal below a critical density of $\rho_c = 10^{-14}$~g~${\rm cm}^{-3}$, and adiabatic above this density.  The speed of sound is $c=2.2 \times 10^4$~cm~${\rm s}^{-1}$.  

The initial magnetic field is uniform along the rotation axis with mass-to-flux ratio 5, or $B_z=163\mu {\rm G}$.  Edge effects with the magnetic field are avoided by embedding the core in an ambient medium in a periodic box of length $4R$.  The medium has a density contrast of 1:30 and is set in pressure equilibrium with the core.

The core is simulated using $1 \times 10^6$ particles.  Self-gravity is included by use of a hierarchical binary tree \citep{benzetal90}, with the SPH smoothing kernel used for gravitational force softening \citep{pm07}.  A sink particle \citep{bbp95} is inserted once the density reaches $\rho_d = 10^{-10}$~g~${\rm cm}^{-3}$ and accretes material within $6.7$ AU.

Only a minimal amount of artificial resistivity is applied to the magnetic field, using the switch described in equation (\ref{eq:intalphaB}) in the range $\alpha_B\in[0,0.1]$.  The constrained hyperbolic divergence cleaning algorithm (\S\ref{sec:cleaning}) is used to remove errors arising from the divergence of the magnetic field.  Using this algorithm, the average divergence error of the field is kept to within $1\%$ (Figure~\ref{fig:jet-cleaning}).

\subsection{First core jet}
\label{sec:jet-results}

\begin{figure}
\centering
 \includegraphics[width=\linewidth]{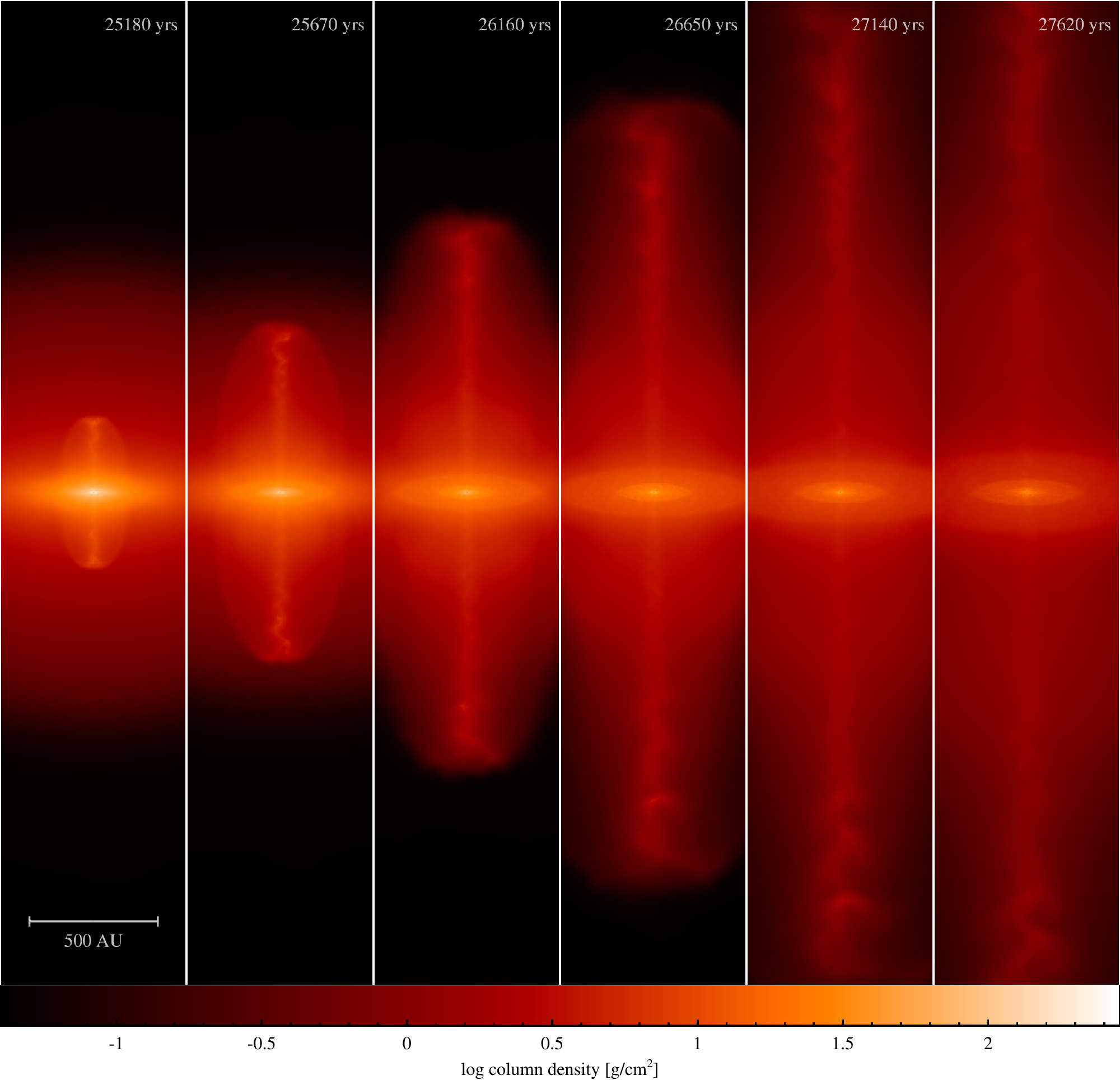}
\caption{Evolution of the jet over its initial 2500 yr period.  It extends several thousand AU during this time.}
\label{fig:jet-jet}
\end{figure}

\begin{figure}
 \includegraphics[width=\linewidth]{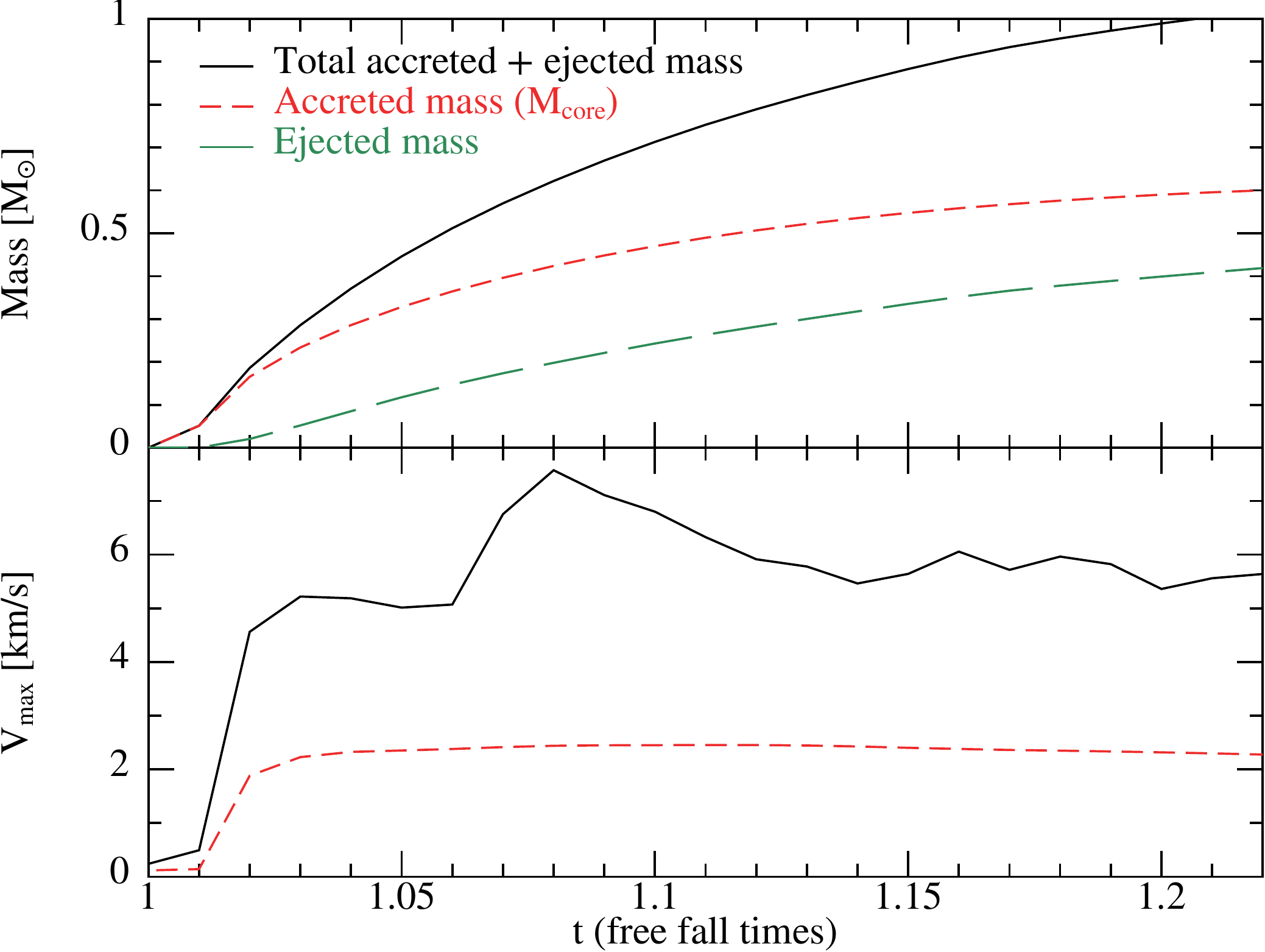}
\caption{The top panel shows the accreted mass onto the sink particle (red, short dashed line) and the mass ejected through the jet (green, long dashed line) over time.  The jet is efficient at removing mass, and continues to do so in our simulation as long as mass is continually supplied.  In the bottom panel, the mean (red, dashed line) and maximum (black, solid line) velocity of the outflow is shown.  The mean velocity is $2$ km ${\rm s}^{-1}$, calculated for particles with speed $>0.1$ km ${\rm s}^{-1}$, with maximum velocities in the range 5--7 km ${\rm s}^{-1}$, consistent with observed outflow velocities.}
\label{fig:jet-massv}
\end{figure}

\begin{figure}
\centering
 \includegraphics[width=0.49\linewidth]{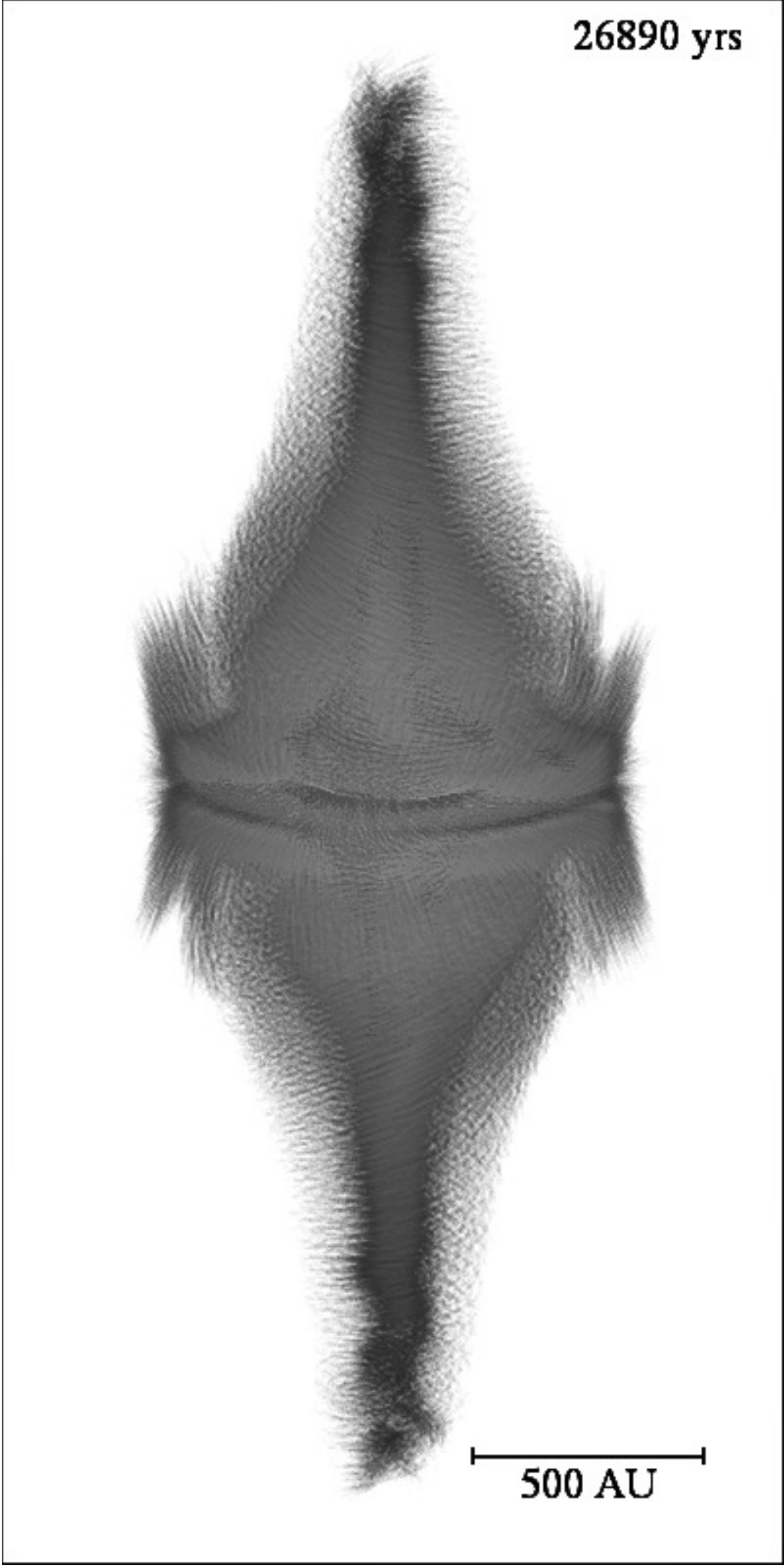} 
 \includegraphics[width=0.49\linewidth]{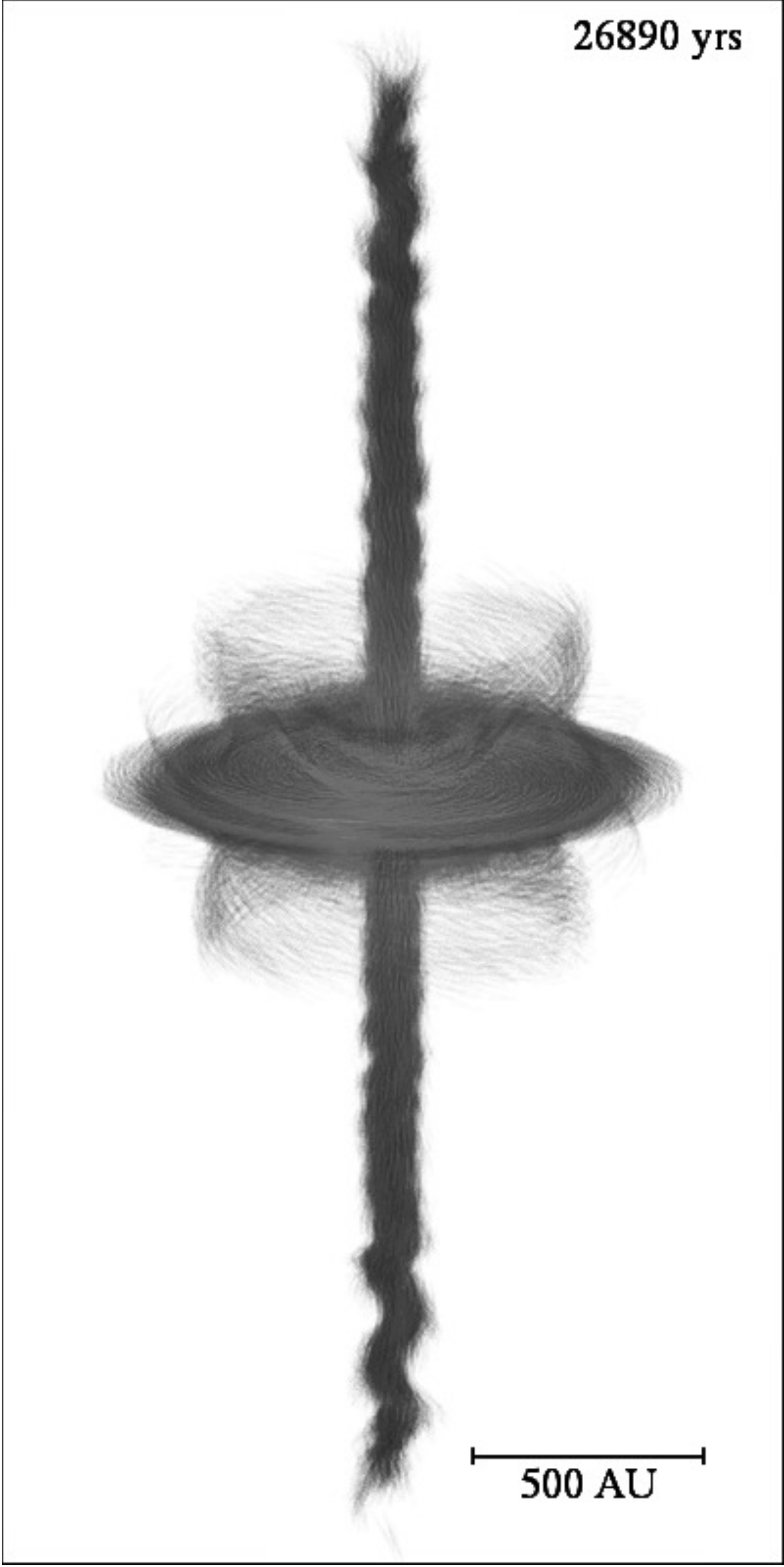} 
\caption{Magnetic field lines (left) and magnetic current (right) at $t_\text{ff}=1.1$.}
\label{fig:jet-field}
\end{figure}

The simulations are performed until $1.4t_\text{ff}$.  At $t_\text{ff}\sim1$, the winding of the magnetic field by the infalling material launches a well collimated jet along the axis of rotation (Figure~\ref{fig:jet-jet}).  The jet has mean velocity $2$~km~${\rm s}^{-1}$, with top end velocities of 5--7~km~${\rm s}^{-1}$ (see Figure~\ref{fig:jet-massv}), which is consistent with observed speeds and the escape velocity for an object of this mass and radius.

The jet is highly efficient at removing mass.  By $t_\text{ff}=1.22$, 40\% of the original material in the core has been ejected through this outflow, consistent with outflow studies by \citet{mm00} and \citet{hansenetal12}.  The material around the sink particle settles into a disc-like object due to the conservation of angular momentum, but has sub-Keplerian orbital velocites by a factor of $\sim$~3--4.

The magnetic field near the sink particle is $\sim$~$100$~mG, wound in a toroidal geometry.  This leads to the ``wiggles'' observed in the jet as a result of the expanding magnetic field in the $z$-direction.  The magnetic field lines at $t_\text{ff}=1.1$ are shown in Figure~\ref{fig:jet-field}, where the field from each particle is represented with an opacity proportional to field strength.

\section{Mach 10 magnetised turbulence}
\label{sec:mhdturb}

\begin{figure*}
\setlength{\tabcolsep}{0.002\textwidth}
 \begin{tabular}{ccccl}
  \includegraphics[height=0.23\linewidth]{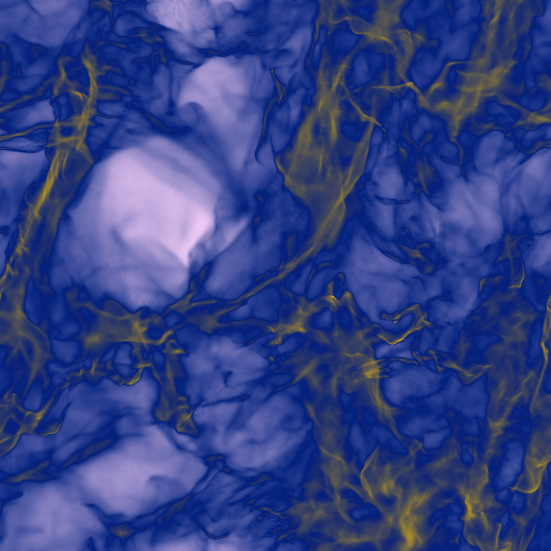} 
& \includegraphics[height=0.23\linewidth]{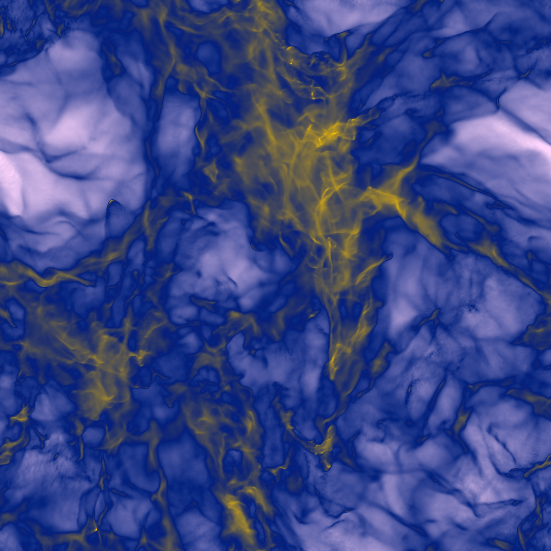} 
& \includegraphics[height=0.23\linewidth]{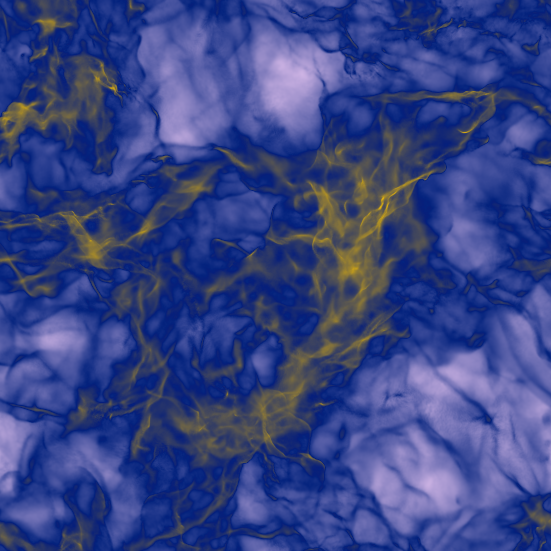} 
& \includegraphics[height=0.23\linewidth]{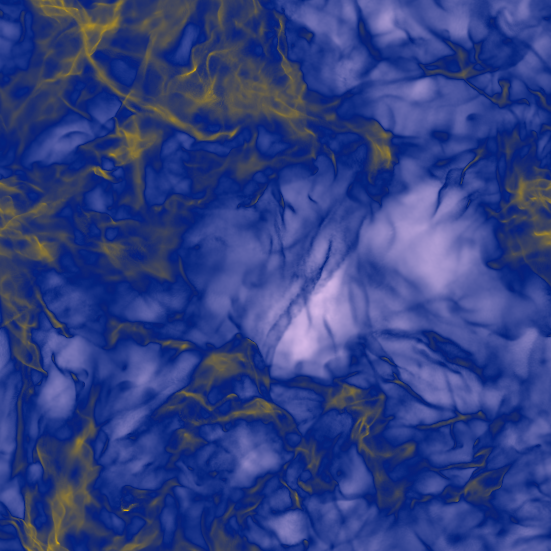} 
& \includegraphics[height=0.23\linewidth]{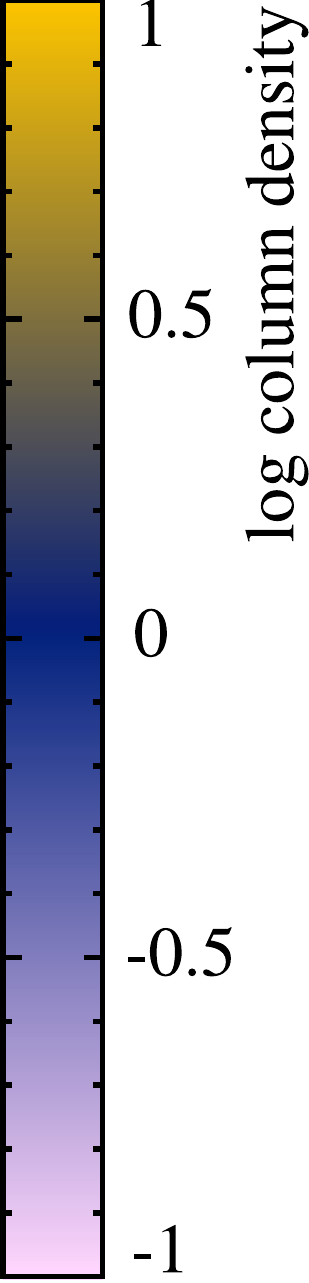} \\
  \includegraphics[height=0.23\linewidth]{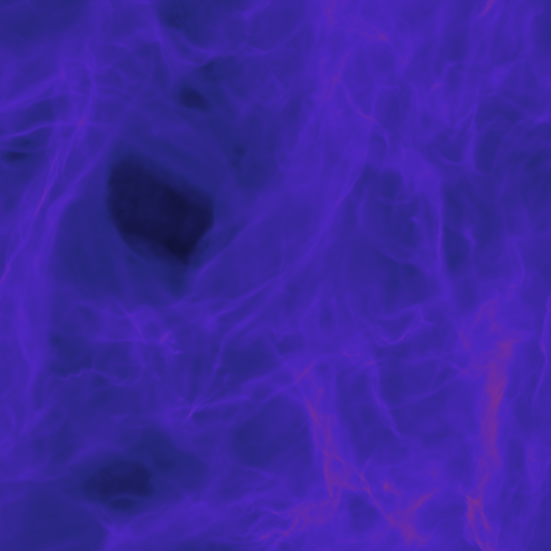} 
& \includegraphics[height=0.23\linewidth]{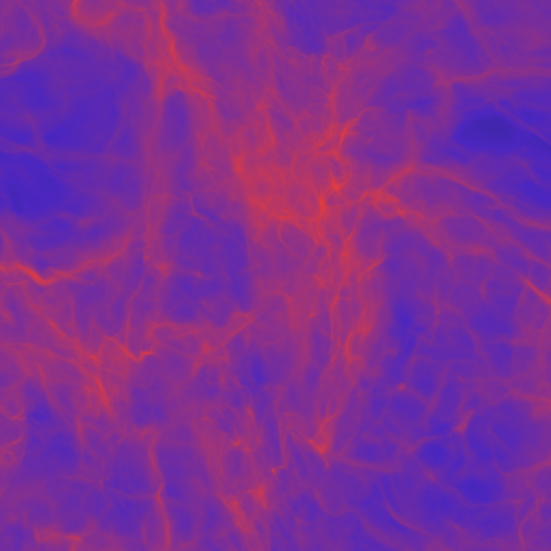} 
& \includegraphics[height=0.23\linewidth]{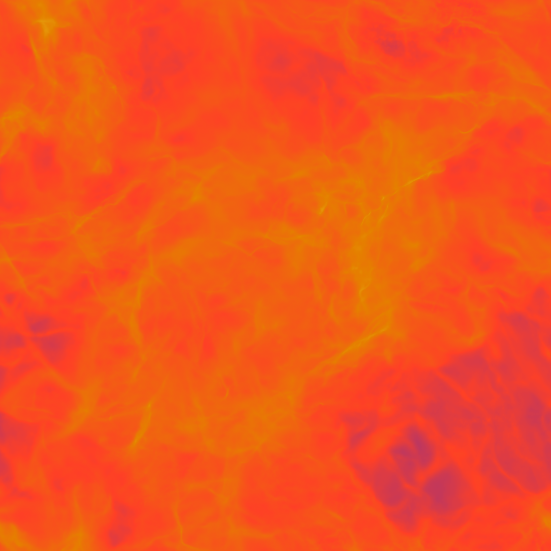} 
& \includegraphics[height=0.23\linewidth]{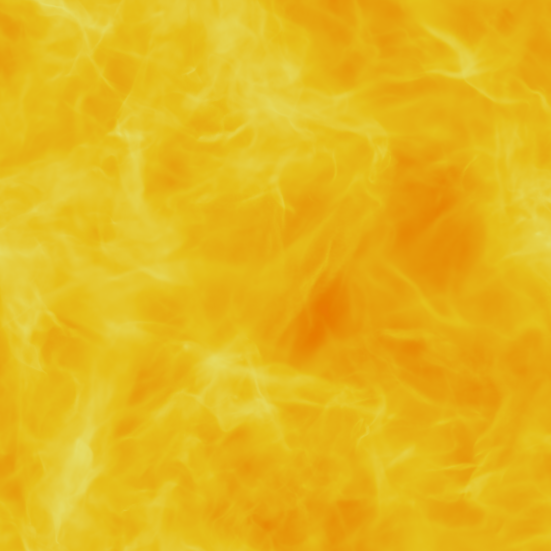} 
& \includegraphics[height=0.23\linewidth]{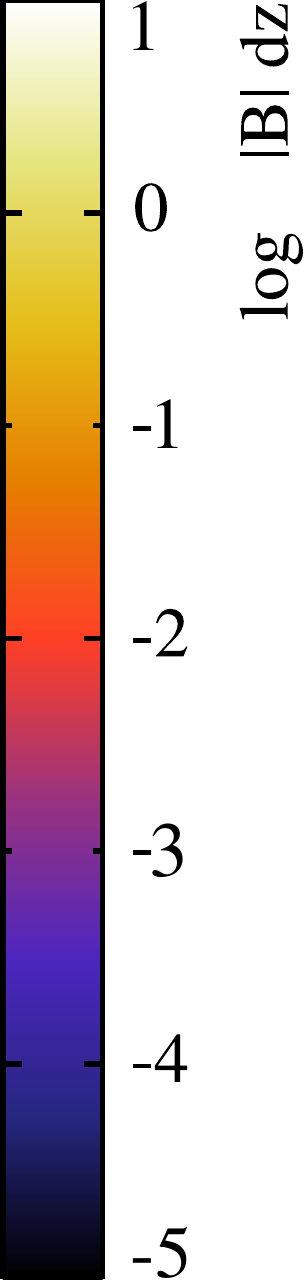}
 \end{tabular}
\caption{Snapshots of the $z$-integrated column density (top) and magnetic field magnitude (bottom) at $t=2,4,8,12$ turbulent turnover times.  $256^3$ SPH particles have been used with the new artificial resistivity switch to reduce magnetic dissipation.}
\label{fig:turb-rendering}
\end{figure*}

Why are observed magnetic fields in the universe as strong as they are?  It is becoming increasingly understood that small scale turbulent dynamos drive exponential amplification of the magnetic field, such that even if initial magnetic fields had tiny field strengths, they would have rapidly reached observed values (see reviews by \citealt{becketal96,bs05,widrowetal12}).

We have been performing a code comparison between SPMHD and finite difference methods on the dynamo amplification of magnetic fields in Mach 10 turbulence, representative of turbulence found in molecular clouds in the Milky Way (for reviews, see \citealt{evans99,es04,mk04,mo07}).  These simulations begin with an initially weak magnetic field, which is exponentially amplified through the conversion of turbulent energy until the magnetic energy reaches equipartition with the kinetic energy (an increase of 10 orders of magnitude in this case).  Our comparison extends the purely hydrodynamic code comparison by \citet{pf10} on the statistics of driven Mach 10 turbulence between SPH and grid-based methods.  Their conclusion was that excellent agreement was found in the properties of the turbulence, with SPH performing better at resolving dense structures, and grid-based methods better for volumetric quantities.

\subsection{Initial conditions}
The simulations are run at resolutions of $128^3$ and $256^3$ particles. The initial conditions are similar to the parameter study performed by \citet{federrathetal11}.  The density is uniform, with $\rho=1$, and an isothermal equation of state is used with $c=1$.  The initial magnetic field is $B_z=\sqrt{2} \times 10^{-5}$ such that the initial plasma beta is $\beta=P_\text{thermal}/P_\text{magnetic}=10^{10}$.

The turbulence is driven and sustained using an acceleration based upon the Ornstein-Uhlenbeck process \citep{eswaranpope88, federrathetal10}, which is a stochastic process with a finite autocorrelation timescale that drives motion at low wave numbers.  The driving force is constructed in Fourier space, allowing it to be decomposed into solenoidal and compressive components and for this case we only use the solenoidal component.

\subsection{Detecting magnetic shocks}

It is important to correctly capture shocks in the magnetic field for these simulations.  However, the dissipation from the added artificial resistivity reduces the magnetic Reynolds number.  Since molecular clouds are known to have high kinetic and magnetic Reynolds numbers ($\sim10^6$--$10^9$), it is important to reduce sources of numerical dissipation.  Thus, a switch is used to ``turn off'' artificial resistivity in regions away from shocks.  We found the switch proposed by \citet{pm05} was not able to detect shocks while the magnetic field was weak, leading to significant noise in the magnetic field and spurious growth rates.  

We have developed a new artificial resistivity switch \citep{tp13} for this work, as described in \S\ref{sec:spmhddiss}.  It sets $\alpha_B = h \vert \nabla {\bf B} \vert / \vert {\bf B} \vert$, which measures the relative degree of discontinuity in the magnetic field.  In this way, it is able to detect and capture shocks throughout the several orders of magnitude change in magnetic field strength.

An important note about dissipation terms in SPMHD is that they are resolution dependent.   The dissipation can be effectively halved by doubling the resolution of the simulation.

\subsection{Turbulence results}

The turbulence is simulated for 60 turbulent turnover times using the {\sc Flash} code \citep{flash,flash3} at $128^3$ grid cells, and with the {\sc Phantom} SPMHD code at $128^3$ and $256^3$ particles.  The SPMHD simulations have been run for both artificial resistivity applied with a fixed $\alpha_B=1$ parameter, and using the new artificial resistivity switch to reduce dissipation.  

Renderings of the evolution of the $z$-integrated column density and magnetic field are shown in Figure~\ref{fig:turb-rendering} for the $256^3$ particle simulation using the new artificial resistivity switch.  The structure in the magnetic field closely resembles the shock structures in the density field, which is to be expected since the magnetic field is weak.  Even once the field is reaching saturation, shocks in the magnetic field are still driven primarily by the forcing though subtle differences start to become apparent.

\begin{figure}
 \includegraphics[width=\linewidth]{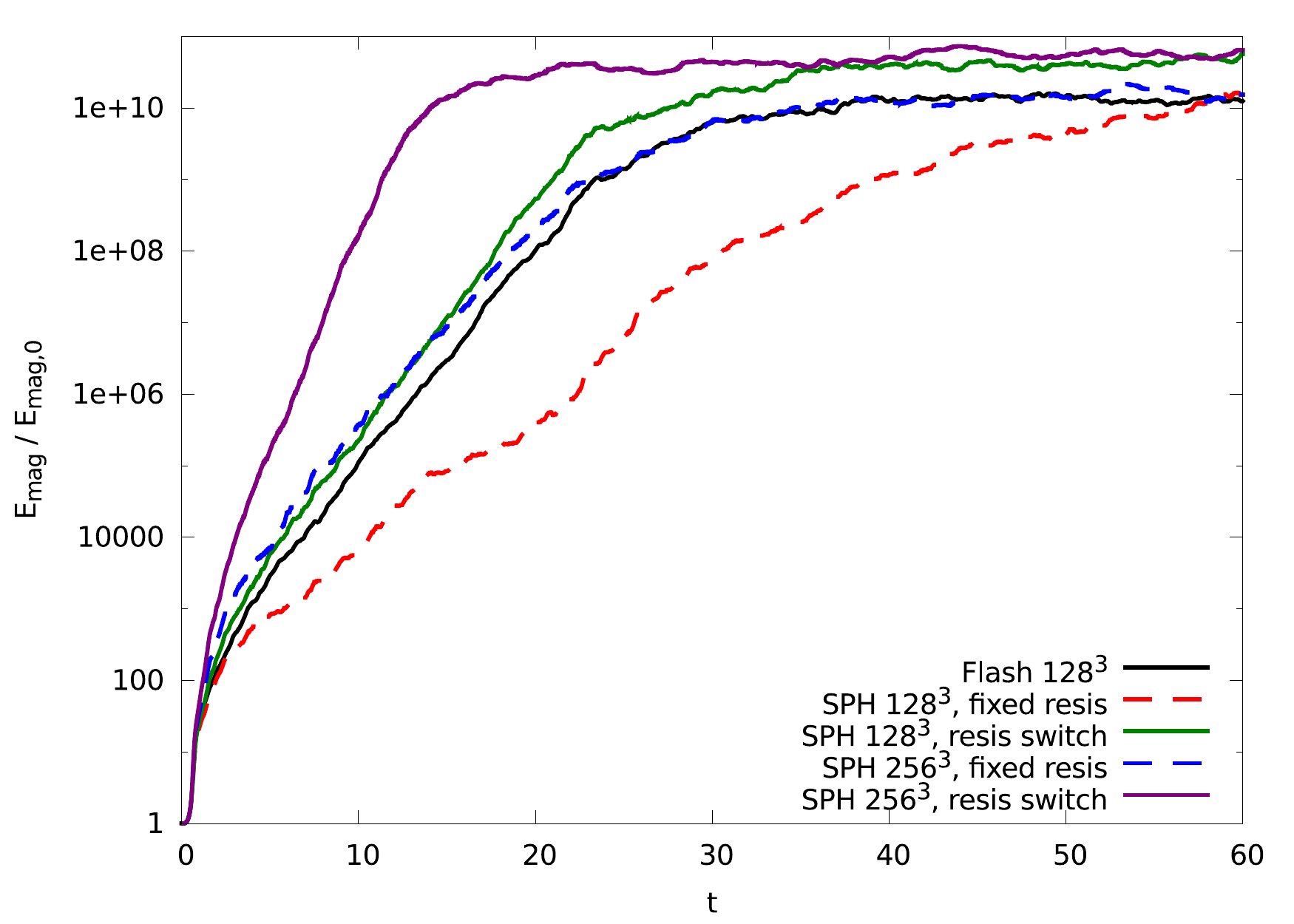}
\caption{Growth of the magnetic energy from turbulent dynamo.  SPH results for $128^3$ and $256^3$ particles using a fixed artificial resistivity parameter and with the new artificial resistivity switch.}
\label{fig:turb-energy}
\end{figure}

The growth of magnetic energy as a function of time is shown in Figure~\ref{fig:turb-energy}.  Comparable growth rates for $128^3$ grid cells can be achieved using either $256^3$ particles when applying fixed artificial resistivity everywhere, or at the same resolution of $128^3$ particles when using the new resistivity switch.  Thus it can be concluded that using the new resistivity switch produces an effect similar to doubling the resolution.

The saturation level is in agreement between the {\sc Flash} results and the SPMHD results when using a fixed artificial resistivity parameter.  When using the new artificial resistivity switch to reduce magnetic dissipation, however, the saturation level for the SPMHD results are 2--3$\times$ higher for both the $128^3$ and $256^3$ simulations.

\section{Summary and discussion}
\label{sec:summary}

In this work, we have performed Smoothed Particle Magnetohydrodynamic (SPMHD) simulations of the gravitational collapse of a prestellar core to form the first hydrostatic core, and also of magnetised Mach 10 turbulence, representative of conditions in molecular clouds.  These simulations used the constrained hyperbolic divergence cleaning method \citep{tp12} to maintain the $\nabla \cdot {\bf B}=0$ constraint on the magnetic field.  It is a Hamiltonian version of hyperbolic divergence cleaning \citep{dedner02} that we have formulated for SPMHD through derivation from the discretised Lagrangian. Thus it possesses the conservation and stability properties inherent to SPH, and was found to reduce errors in the magnetic field by 10$\times$. 

The collapsing core simulations \citep{ptb12} produced a slow, well collimated jet from the central object that has properties consistent with candidate observations of first hydrostatic cores.  It is efficient at removing mass out of the core, with up to 40\% of the material being removed by the time the remaining mass in the core has been accreted onto the central sink particle.

Our simulations of Mach 10 magnetised turbulence model dynamo amplification of the magnetic field through the conversion of turbulent energy into magnetic energy.  An initially weak magnetic field is present which was exponentially increased $\sim$~10 orders of magnitude in energy until it reaches saturation.  Both the saturation level and growth rates were consistent with results from the grid based code {\sc Flash} when using a fixed artificial resistivity parameter.  Simulations were also run using a switch for artificial resistivity to reduce numerical dissipation \citep{tp13}, which lead to similar growth rates as if double the resolution had been used, but with a saturation level 2--3$\times$ higher. \newline

T. Tricco thanks the conference organisers for the opportunity to speak.  T. Tricco is supported by Endeavour IPRS and APA postgraduate research scholarships.  We are grateful for funding via Australian Research Council Discovery Projects grants DP1094585 and DP110102191.  This research was undertaken with the assistance of resources provided at the Multi-modal Australian ScienceS Imaging and Visualisation Environment (MASSIVE) through the National Computational Merit Allocation Scheme supported by the Australian Government. The {\sc Flash} simulations were run at LRZ (grant pr32lo) and JSC (grant hhd20).

\bibliographystyle{rmaa}
\bibliography{terrence-tricco--mfu4-bib}

\begin{thebibliography}
\expandafter\ifx\csname natexlab\endcsname\relax\def\natexlab#1{#1}\fi
\expandafter\ifx\csname href\endcsname\relax
  \def\href#1#2{}\fi
\expandafter\ifx\csname urllinklabel\endcsname\relax
  \def\urllinklabel{[LINK]}\fi
\expandafter\ifx\csname adsurllinklabel\endcsname\relax
  \def\adsurllinklabel{[ADS]}\fi

\bibitem[{{Bate} {et~al.}(1995){Bate}, {Bonnell}, \& {Price}}]{bbp95}
{Bate}, M.~R., {Bonnell}, I.~A., \& {Price}, N.~M. 1995, MNRAS, 277, 362


\bibitem[{{Bate} \& {Burkert}(1997)}]{bb97}
{Bate}, M.~R. \& {Burkert}, A. 1997, MNRAS, 288, 1060


\bibitem[{{Beck} {et~al.}(1996){Beck}, {Brandenburg}, {Moss}, {Shukurov}, \&
  {Sokoloff}}]{becketal96}
{Beck}, R., {Brandenburg}, A., {Moss}, D., {Shukurov}, A., \& {Sokoloff}, D.
  1996, ARA\&A, 34, 155


\bibitem[{{Benz} {et~al.}(1990){Benz}, {Cameron}, {Press}, \&
  {Bowers}}]{benzetal90}
{Benz}, W., {Cameron}, A.~G.~W., {Press}, W.~H., \& {Bowers}, R.~L. 1990, ApJ,
  348, 647


\bibitem[{{Brandenburg} \& {Subramanian}(2005)}]{bs05}
{Brandenburg}, A. \& {Subramanian}, K. 2005, Phys. Rep., 417, 1


\bibitem[{{Carroll} {et~al.}(2010){Carroll}, {Frank}, \& {Blackman}}]{cfb10}
{Carroll}, J.~J., {Frank}, A., \& {Blackman}, E.~G. 2010, ApJ, 722, 145


\bibitem[{{Commer{\c c}on} {et~al.}(2010){Commer{\c c}on}, {Hennebelle},
  {Audit}, {Chabrier}, \& {Teyssier}}]{commerconetal10}
{Commer{\c c}on}, B., {Hennebelle}, P., {Audit}, E., {Chabrier}, G., \&
  {Teyssier}, R. 2010, A\&A, 510, L3


\bibitem[{{Dedner} {et~al.}(2002){Dedner}, {Kemm}, {Kr{\"o}ner}, {Munz},
  {Schnitzer}, \& {Wesenberg}}]{dedner02}
{Dedner}, A., {Kemm}, F., {Kr{\"o}ner}, D., {Munz}, C.-D., {Schnitzer}, T., \&
  {Wesenberg}, M. 2002, J. Comput. Phys., 175, 645


\bibitem[{{Dubey} {et~al.}(2008){Dubey}, {Fisher}, {Graziani}, {Jordan},
  {Lamb}, {Reid}, {Rich}, {Sheeler}, {Townsley}, \& {Weide}}]{flash3}
{Dubey}, A., {Fisher}, R., {Graziani}, C., {Jordan}, IV, G.~C., {Lamb}, D.~Q.,
  {Reid}, L.~B., {Rich}, P., {Sheeler}, D., {Townsley}, D., \& {Weide}, K.
  Astronomical Society of the Pacific Conference Series, Vol. 385, , Numerical
  Modeling of Space Plasma Flows, ed. N.~V. {Pogorelov}E.~{Audit} \& G.~P.
  {Zank}, 145


\bibitem[{{Dunham} {et~al.}(2011){Dunham}, {Chen}, {Arce}, {Bourke}, {Schnee},
  \& {Enoch}}]{dunhametal11}
{Dunham}, M.~M., {Chen}, X., {Arce}, H.~G., {Bourke}, T.~L., {Schnee}, S., \&
  {Enoch}, M.~L. 2011, ApJ, 742, 1


\bibitem[{{Elmegreen} \& {Scalo}(2004)}]{es04}
{Elmegreen}, B.~G. \& {Scalo}, J. 2004, ARA\&A, 42, 211


\bibitem[{{Enoch} {et~al.}(2010){Enoch}, {Lee}, {Harvey}, {Dunham}, \&
  {Schnee}}]{enochetal10}
{Enoch}, M.~L., {Lee}, J.-E., {Harvey}, P., {Dunham}, M.~M., \& {Schnee}, S.
  2010, ApJL, 722, L33


\bibitem[{{Eswaran} \& {Pope}(1988)}]{eswaranpope88}
{Eswaran}, V. \& {Pope}, S.~B. 1988, Computers and Fluids, 16, 257


\bibitem[{{Evans}(1999)}]{evans99}
{Evans}, II, N.~J. 1999, ARA\&A, 37, 311


\bibitem[{{Federrath} {et~al.}(2011){Federrath}, {Chabrier}, {Schober},
  {Banerjee}, {Klessen}, \& {Schleicher}}]{federrathetal11}
{Federrath}, C., {Chabrier}, G., {Schober}, J., {Banerjee}, R., {Klessen},
  R.~S., \& {Schleicher}, D.~R.~G. 2011, Physical Review Letters, 107, 114504


\bibitem[{{Federrath} \& {Klessen}(2012)}]{fk12}
{Federrath}, C. \& {Klessen}, R.~S. 2012, ApJ, 761, 156


\bibitem[{{Federrath} {et~al.}(2010){Federrath}, {Roman-Duval}, {Klessen},
  {Schmidt}, \& {Mac Low}}]{federrathetal10}
{Federrath}, C., {Roman-Duval}, J., {Klessen}, R.~S., {Schmidt}, W., \& {Mac
  Low}, M.-M. 2010, A\&A, 512, A81


\bibitem[{{Fryxell} {et~al.}(2000){Fryxell}, {Olson}, {Ricker}, {Timmes},
  {Zingale}, {Lamb}, {MacNeice}, {Rosner}, {Truran}, \& {Tufo}}]{flash}
{Fryxell}, B., {Olson}, K., {Ricker}, P., {Timmes}, F.~X., {Zingale}, M.,
  {Lamb}, D.~Q., {MacNeice}, P., {Rosner}, R., {Truran}, J.~W., \& {Tufo}, H.
  2000, ApJS, 131, 273


\bibitem[{{Hansen} {et~al.}(2012){Hansen}, {Klein}, {McKee}, \&
  {Fisher}}]{hansenetal12}
{Hansen}, C.~E., {Klein}, R.~I., {McKee}, C.~F., \& {Fisher}, R.~T. 2012, ApJ,
  747, 22


\bibitem[{{Larson}(1969)}]{larson69}
{Larson}, R.~B. 1969, MNRAS, 145, 271


\bibitem[{{Mac Low} \& {Klessen}(2004)}]{mk04}
{Mac Low}, M.-M. \& {Klessen}, R.~S. 2004, Reviews of Modern Physics, 76, 125


\bibitem[{{Machida} \& {Hosokawa}(2013)}]{mh13}
{Machida}, M.~N. \& {Hosokawa}, T. 2013, MNRAS, 431, 1719


\bibitem[{{Machida} {et~al.}(2008){Machida}, {Inutsuka}, \&
  {Matsumoto}}]{mim08}
{Machida}, M.~N., {Inutsuka}, S.-i., \& {Matsumoto}, T. 2008, ApJ, 676, 1088


\bibitem[{{Matzner} \& {McKee}(2000)}]{mm00}
{Matzner}, C.~D. \& {McKee}, C.~F. 2000, ApJ, 545, 364


\bibitem[{{McKee} \& {Ostriker}(2007)}]{mo07}
{McKee}, C.~F. \& {Ostriker}, E.~C. 2007, ARA\&A, 45, 565


\bibitem[{{Monaghan}(1997)}]{monaghan97}
{Monaghan}, J.~J. 1997, J. Comput. Phys., 136, 298


\bibitem[{{Monaghan}(2005)}]{monaghan05}
---. 2005, Reports on Progress in Physics, 68, 1703


\bibitem[{{Monaghan} \& {Lattanzio}(1985)}]{ml85}
{Monaghan}, J.~J. \& {Lattanzio}, J.~C. 1985, A\&A, 149, 135


\bibitem[{{Morris} \& {Monaghan}(1997)}]{mm97}
{Morris}, J.~P. \& {Monaghan}, J.~J. 1997, J. Comput. Phys., 136, 41


\bibitem[{{Nakamura} \& {Li}(2007)}]{nl07}
{Nakamura}, F. \& {Li}, Z.-Y. 2007, ApJ, 662, 395


\bibitem[{{Padoan} \& {Nordlund}(2011)}]{pn11}
{Padoan}, P. \& {Nordlund}, {\AA}. 2011, ApJ, 730, 40


\bibitem[{{Pineda} {et~al.}(2011){Pineda}, {Arce}, {Schnee}, {Goodman},
  {Bourke}, {Foster}, {Robitaille}, {Tanner}, {Kauffmann}, {Tafalla},
  {Caselli}, \& {Anglada}}]{pinedaetal11}
{Pineda}, J.~E., {Arce}, H.~G., {Schnee}, S., {Goodman}, A.~A., {Bourke}, T.,
  {Foster}, J.~B., {Robitaille}, T., {Tanner}, J., {Kauffmann}, J., {Tafalla},
  M., {Caselli}, P., \& {Anglada}, G. 2011, ApJ, 743, 201


\bibitem[{{Price}(2012)}]{price12}
{Price}, D.~J. 2012, J. Comput. Phys., 231, 759


\bibitem[{{Price} \& {Federrath}(2010)}]{pf10}
{Price}, D.~J. \& {Federrath}, C. 2010, MNRAS, 406, 1659


\bibitem[{{Price} \& {Monaghan}(2004{\natexlab{a}})}]{pm04a}
{Price}, D.~J. \& {Monaghan}, J.~J. 2004{\natexlab{a}}, MNRAS, 348, 123


\bibitem[{{Price} \& {Monaghan}(2004{\natexlab{b}})}]{pm04b}
---. 2004{\natexlab{b}}, MNRAS, 348, 139


\bibitem[{{Price} \& {Monaghan}(2005)}]{pm05}
---. 2005, MNRAS, 364, 384


\bibitem[{{Price} \& {Monaghan}(2007)}]{pm07}
---. 2007, MNRAS, 374, 1347


\bibitem[{{Price} {et~al.}(2012){Price}, {Tricco}, \& {Bate}}]{ptb12}
{Price}, D.~J., {Tricco}, T.~S., \& {Bate}, M.~R. 2012, MNRAS, 423, L45


\bibitem[{{Richer} {et~al.}(2000){Richer}, {Shepherd}, {Cabrit}, {Bachiller},
  \& {Churchwell}}]{richeretal00}
{Richer}, J.~S., {Shepherd}, D.~S., {Cabrit}, S., {Bachiller}, R., \&
  {Churchwell}, E. 2000, Protostars and Planets IV, 867


\bibitem[{{Springel} \& {Hernquist}(2002)}]{sh02}
{Springel}, V. \& {Hernquist}, L. 2002, MNRAS, 333, 649


\bibitem[{{Tricco} \& {Price}(2012)}]{tp12}
{Tricco}, T.~S. \& {Price}, D.~J. 2012, J. Comput. Phys., 231, 7214


\bibitem[{{Tricco} \& {Price}(submitted)}]{tp13}
---. submitted, MNRAS


\bibitem[{{Widrow} {et~al.}(2012){Widrow}, {Ryu}, {Schleicher}, {Subramanian},
  {Tsagas}, \& {Treumann}}]{widrowetal12}
{Widrow}, L.~M., {Ryu}, D., {Schleicher}, D.~R.~G., {Subramanian}, K.,
  {Tsagas}, C.~G., \& {Treumann}, R.~A. 2012, Space Sci. Rev., 166, 37


\bibitem[{{Wu} {et~al.}(2004){Wu}, {Wei}, {Zhao}, {Shi}, {Yu}, {Qin}, \&
  {Huang}}]{wuetal04}
{Wu}, Y., {Wei}, Y., {Zhao}, M., {Shi}, Y., {Yu}, W., {Qin}, S., \& {Huang}, M.
  2004, A\&A, 426, 503


\end{thebibliography}

\end{document}